\documentclass[epj-spec]{svjour}
\usepackage{graphicx, wrapfig}
\begin{document}
\title{COMPASS Calorimetry in view of future plans}
\author{F.~Nerling for the COMPASS Collaboration}
%\offprints{F.~Nerling}          % Insert a name or remove this line
\mail{Frank.Nerling@cern.ch}
\institute{Universit\"at Freiburg, Physikalisches Institut, Freiburg, Germany }
\abstract{
The COMPASS experiment at the CERN SPS is dedicated to hadron physics with a 
broad research programme, including the study of the nucleon spin structure using muons as a 
probe and a variety of issues in meson spectroscopy using hadron beams. The two stage fixed 
target spectrometer with electromagnetic (em) and hadronic calorimetry in both stages provides 
photon detection in a wide angular range. As discussed in this paper, the COMPASS em calorimetry plays 
a crucial r\^{o}le for the Hadron programme started in 2008 as well as for the planned COMPASS future 
programme of measuring GPDs via exclusive DVCS photons. 
We present the photon detection coverage foreseen, and first, preliminary results characterising 
the present performances of both existing COMPASS electromagnetic calorimeters, based 
on test beam data taken at CERN T9 facility end of 2007.
} 
%end of abstract
\maketitle
\section{Introduction \& motivation}
\label{intro}
\vspace{-0.2cm}
%-------------------------------------------------------
%
%   Section: Intro & motivation
%
%------------------------------------------------------
\subsection{The COMPASS experiment}
\label{subsec.1}
\vspace{-0.2cm}
The COMPASS fixed target experiment~\cite{spectro} at CERN SPS is dedicated to the study of nucleon spin structure
and hadron spectroscopy, addressing the question of how nucleons and hadrons in general are built up from
quarks and gluons~\cite{compass}.
The COMPASS Collaboration has already collected data scattering polarised muon beam of 160\,GeV/c 
on polarised deuteron ($^{6}$LiD) and proton (NH$_{3}$) targets during the years 2002-2004 and 2006-2007. 
The gluon contribution to the nucleon spin is one example of physics determined from these data. 
For the hadron programme, merely a pilot run was performed in 2004, focusing on measuring the Primakoff
reaction on a Pb target. Also some diffractive pion data on that target had been taken and pion dissociation 
into $\pi^{-}\pi^{-}\pi^{+}$ have been analysed~\cite{quirinPraha08}.

\begin{figure}[t]
\noindent
\centering
\includegraphics[clip, trim= 40 50 45 90,width=0.8\linewidth]{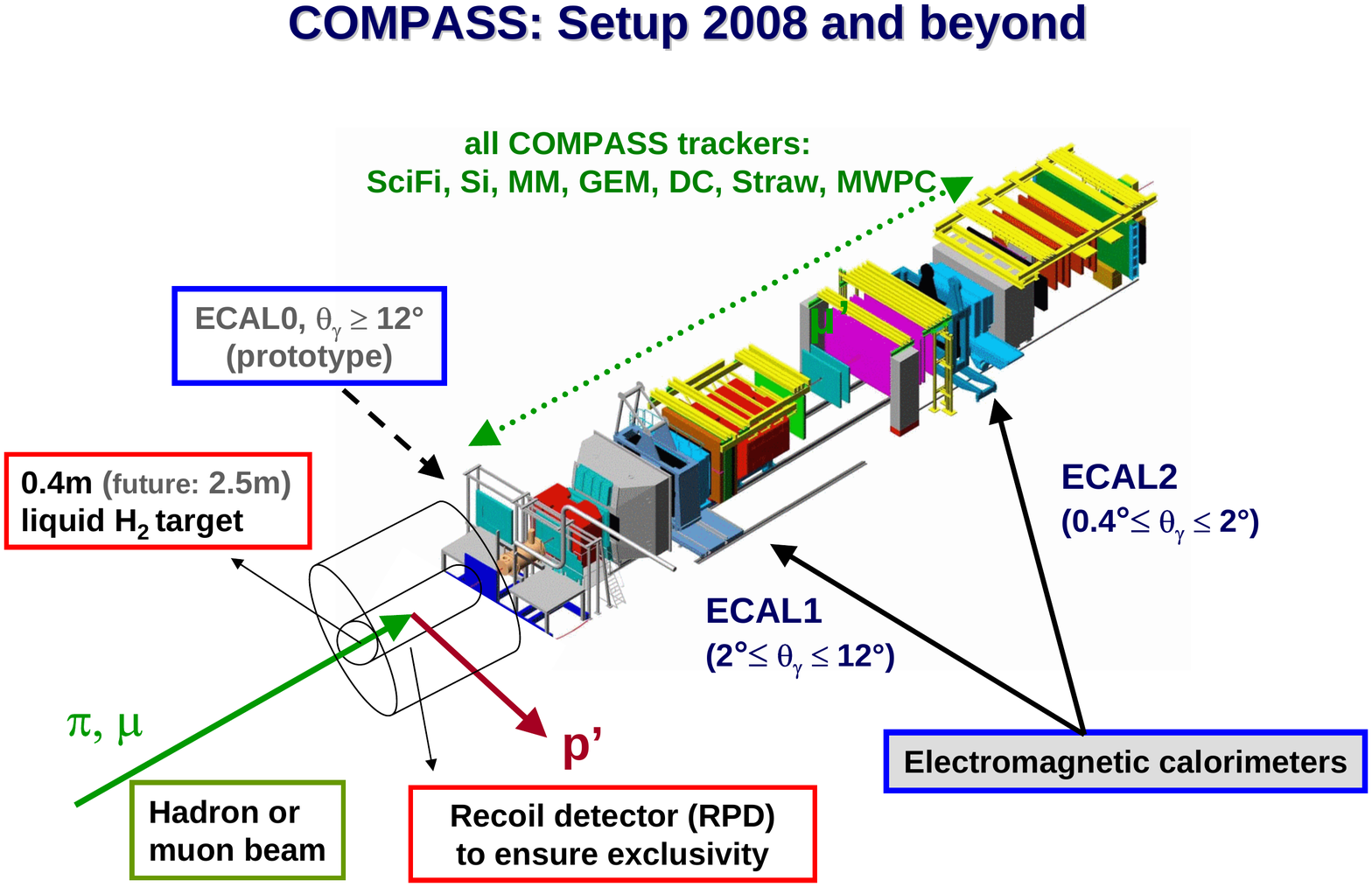}
\caption{\emph{Sketch of the COMPASS experimental setup as used in 2008 and beyond - not to scale.}}
\label{fig.setup2008}
\end{figure}

In 2008 we have started to take high statistics data for spectroscopy of the light hadron sector at high 
energy (190\,GeV/c, $\pi^{-}$ beam). Pion-proton reactions comprising both diffractive and centrally produced
final states allow the search for $J^{PC}$-exotic mesons, glueballs and hybrids. A sketch of the COMPASS
spectrometer as used in 2008 is shown in Fig.\ref{fig.setup2008}, for details see \cite{spectro}. 
Emphasised are the existing em-calorimeters ECAL1, ECAL2 as well as ECAL0, which is foreseen to detect DVCS 
photons under angles larger than 12 degree, cf. Sec.\,\ref{subsec.2}. In 2008, a 40\,cm long liquid hydrogen 
target (LH2) is used. To ensure the exclusivity of the measurements, it is surrounded by a newly introduced 
Recoil Proton Detector (RPD) performing a time-of-flight measurement of recoiling particles like protons.
It consists of two concentric barrels of scintillator counters, read-out at both sides. It provides
particle-identification capability and measures the recoiling proton momentum at few percent accuracy.
\begin{figure}[b]
\vspace{-0.7cm}
\begin{minipage}[c]{.42\linewidth}
\begin{center}
\includegraphics[clip, trim= 250 180 270 195, width=1.\linewidth]{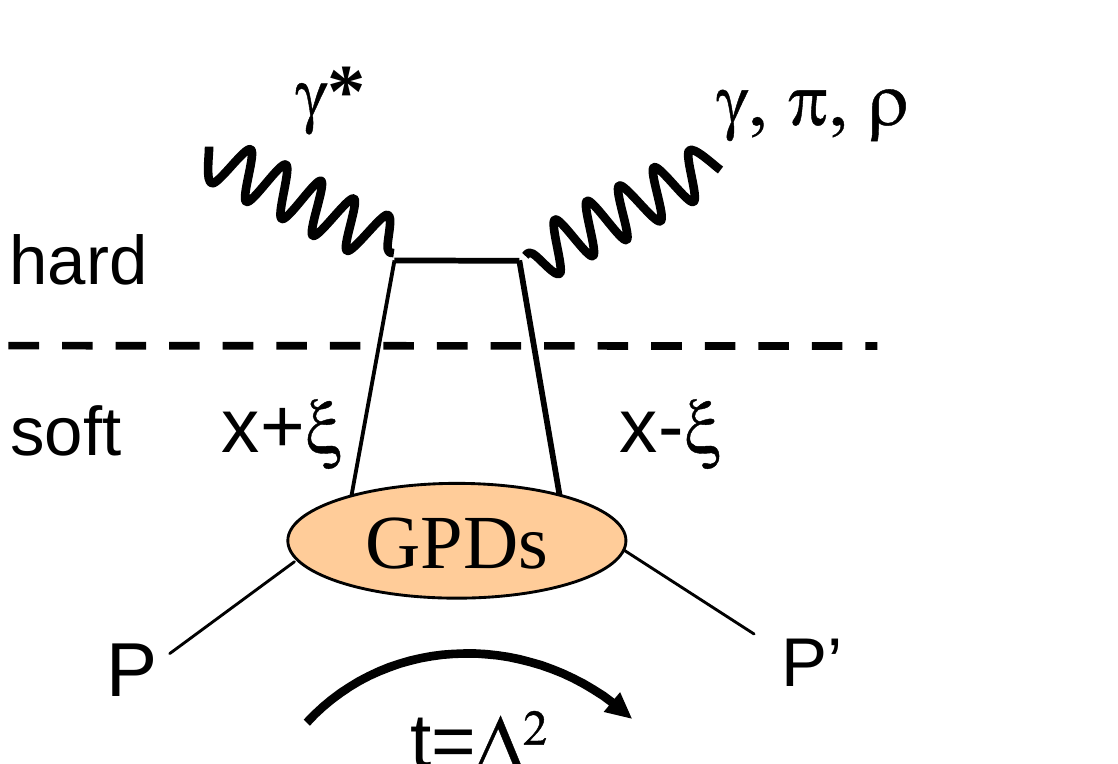}
\end{center}
\caption{\emph{Handbag diagram for DVCS and HEMP reactions: longitudinal quark momentum fraction $x$, 
longitudinal momentum transfer $\xi=x_{\rm Bj}/(2-x_{\rm Bj})$ to the nucleon, and the momentum transfer 
squared $t$ to the target nucleon.}}
\label{fig.dvcs-hemp}
\end{minipage}\hfill
\begin{minipage}[c]{.54\linewidth}
\begin{center}
\vspace{-0.9cm}
\includegraphics[clip, trim= 230 72 220 313,width=1.\linewidth]{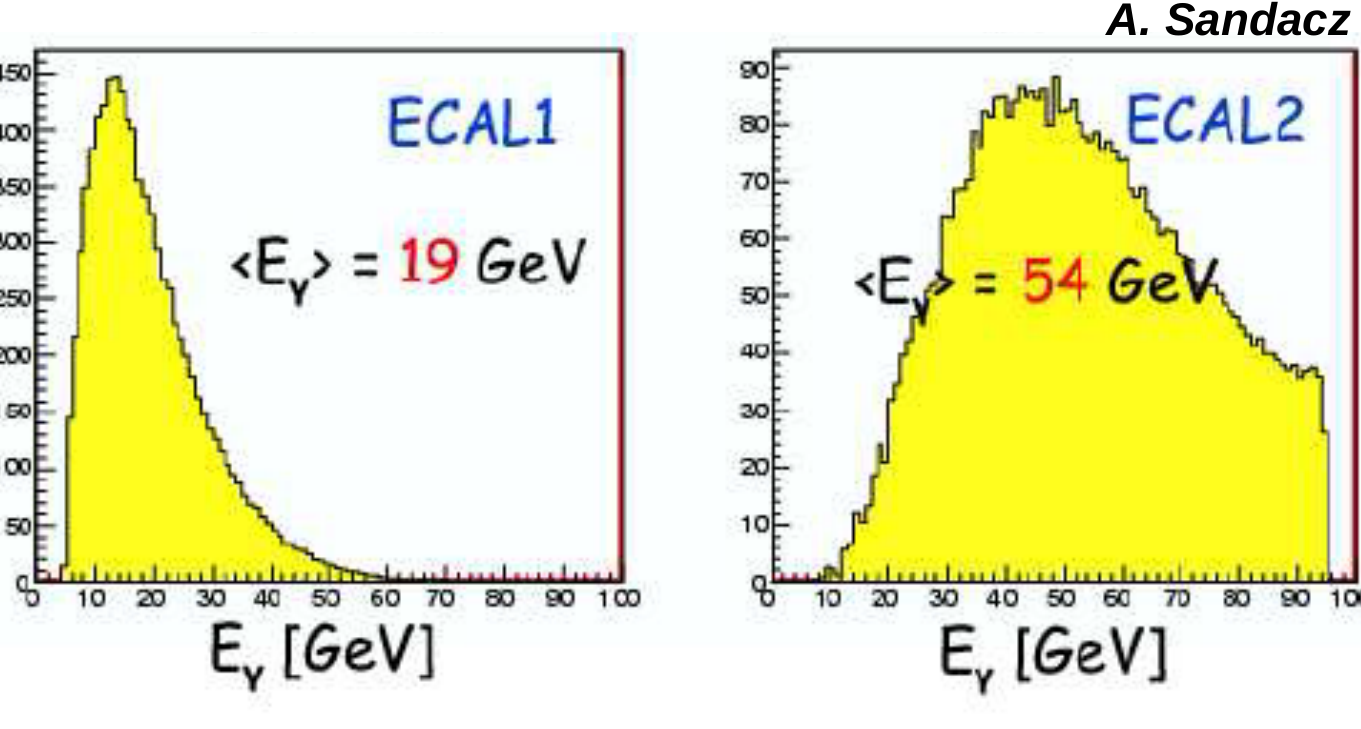}
\end{center}
\caption{\emph{Simulated energy distribution of DVCS photons to be detected in ECAL1 and ECAL2. 
The minimum energies of DVCS photons to be detected in ECAL1 and ECAL2 are 5\,GeV and 10\,GeV respectively.}}
\label{fig.dvcsEdistri}
\end{minipage}
\end{figure}
\subsection{Future plans}
\label{subsec.2}
\vspace{-0.25cm}
The COMPASS Collaboration has expressed the interest for pursuing an experiment dedicated to the 
measurement of Generalised Parton Distributions (GPDs)~\cite{eoi_gpd}. This novel formalism
provides a three dimensional picture of partons inside the nucleon (longitudinal momentum fraction $x$ and 
transverse impact parameter). In addition, the second moment of GPDs gives access to the total angular 
momentum carried by the partons inside the nucleon via the Ji sum rule~\cite{jiSumRule}, and hence they can
provide new insights on the nucleon spin puzzle. The GPDs are accessible via hard exclusive reactions like 
Deep Virtual Compton Scattering (DVCS) and Hard Exclusive Meson Production (HEMP), see Fig.\,\ref{fig.dvcs-hemp}. 
The plan is to measure them on (liquid) proton and deuterium targets. 

As for the Hadron programme, the RPD detector 
is crucial to ensure exclusivity for measurements of DVCS and HEMP, for the former the calorimetry plays a mandatory 
r\^{o}le in addition. 
Even though the future GPD measurements are planned on a longer LH2 target, namely the order of a few meters, 
and ECAL0 is not yet designed, the ideal synergy and complementarity of both programmes is obvious. Since no big change over 
of the spectrometer is needed, both hadron and muon beams are available at CERN/SPS and easily switchable, 
we will go for a GPD pilot run during the 2009 Hadron run to study and optimise the feasibility for GPD via DVCS 
measurements.
\subsection{The r\^{o}le of COMPASS calorimetry}
\begin{figure}[t]
\begin{minipage}[c]{.58\linewidth}
\begin{center} 
\includegraphics[clip, trim= 45 65 40 150,width=1.0\linewidth]{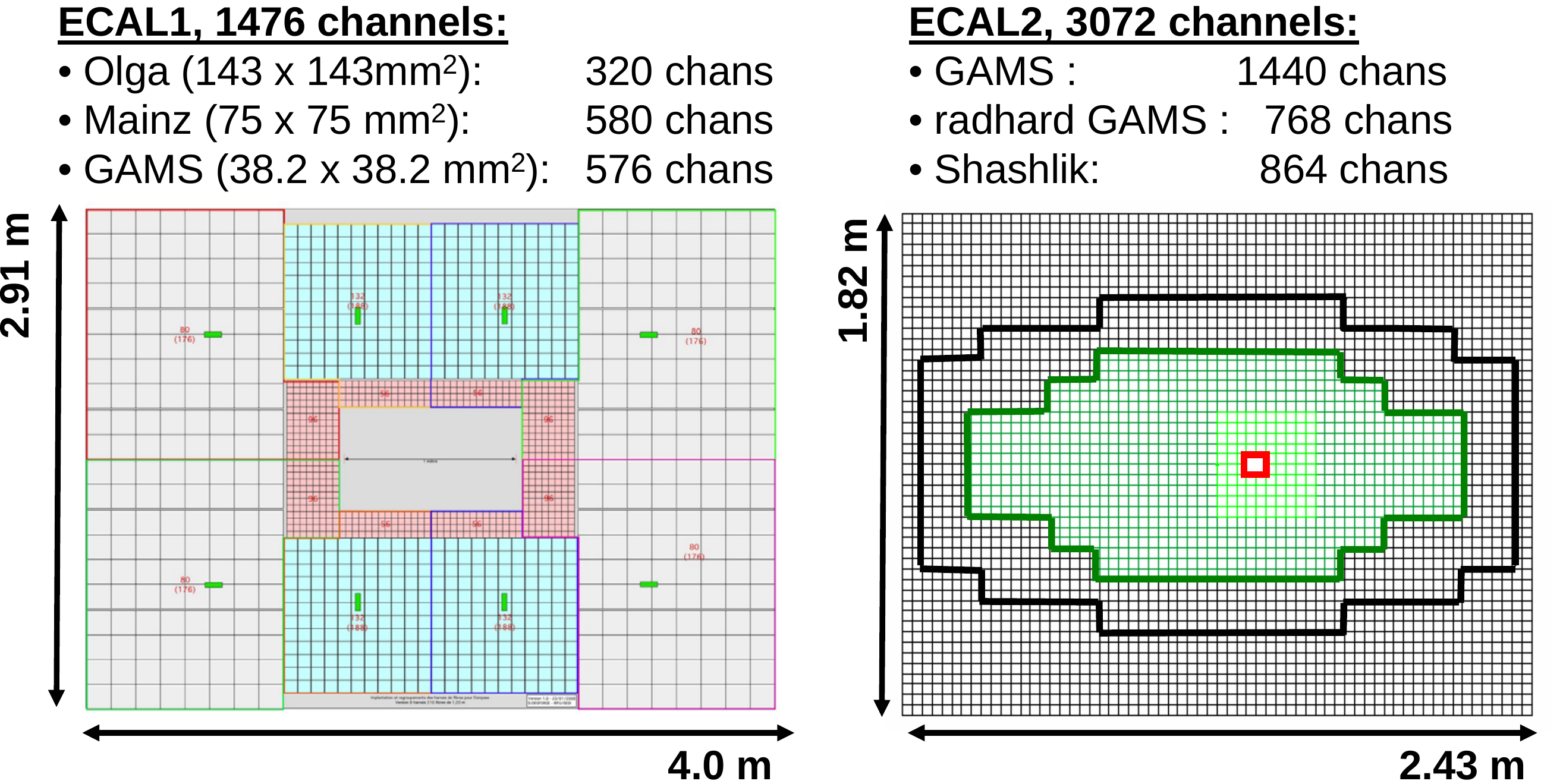}
\end{center}
\caption{\emph{Scheme of existing em-calorimeters - {\rm (left)} ECAL1: Homogene lead glass Cherenkov counters of different cross
sections - {\rm (right)} ECAL2: New Shashlik sampling modules in central region, radiation hard lead glass (between green and black
border), and GAMS lead glass blocks (outer region same as in ECAL1).}}
\label{fig.ecals1u2}
\end{minipage}\hfill
\begin{minipage}[c]{.38\linewidth}
\begin{center}
\includegraphics[clip, trim= 80 120 300 130, width=1.0\linewidth]{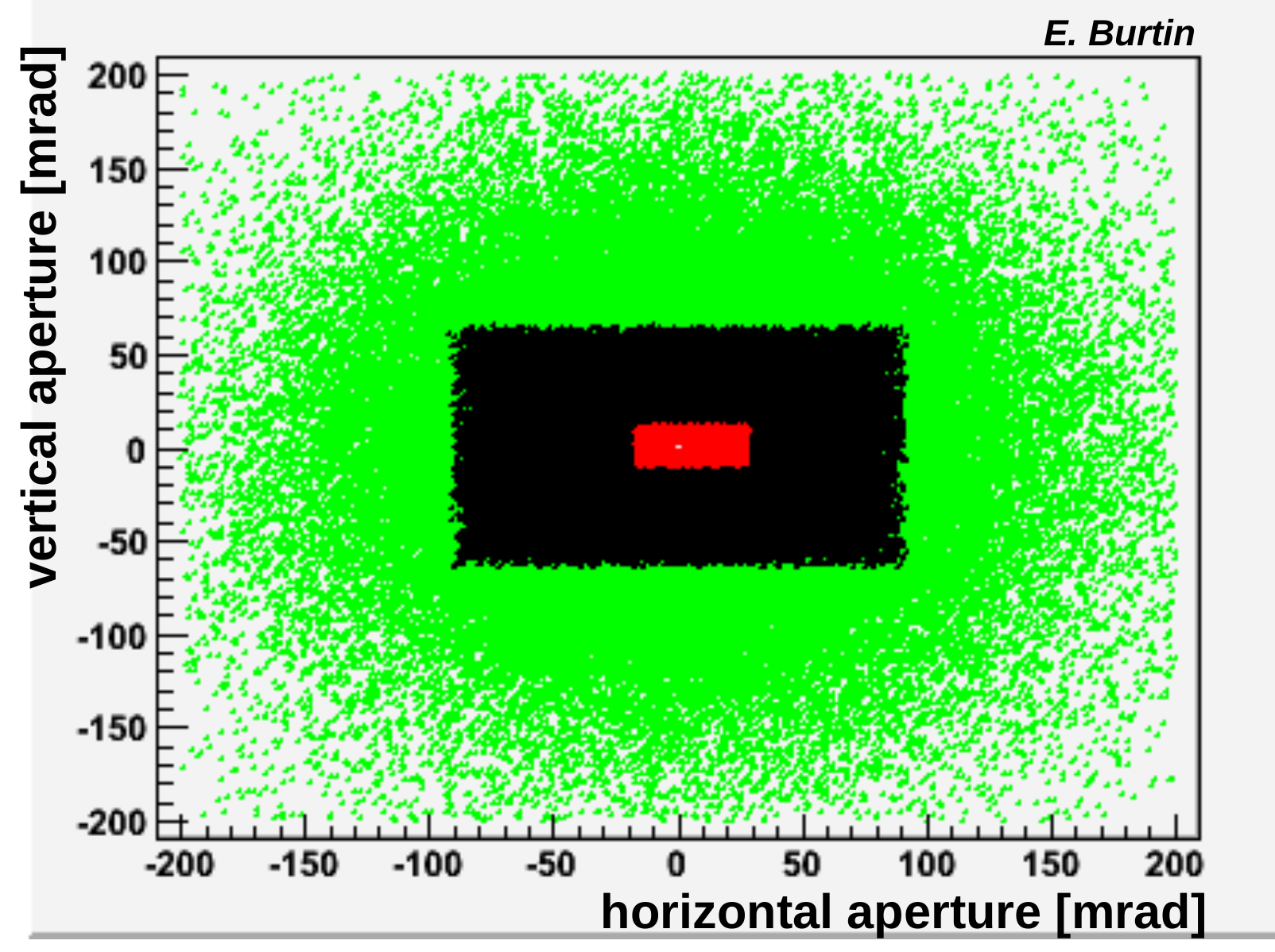}
\end{center}
\caption{\emph{Calorimeter coverage foreseen: Simulated DVCS $\gamma$ impact point at location of ECAL0. 
(ECAL0-green at 2.5\,m downstream of the target, to be built; ECAL1-black at 11.1\,m and ECAL2-red at 
33.25\,m, both existing)}}
\label{fig.ecalCoverage}
\end{minipage}
\end{figure}

\vspace{-0.25cm}
The em-calorimeters play a crucial r\^{o}le for both the hadron programme as well as the GPD measurements
via DVCS, since neutral particle detection over a wide angular range are mandatory. 
On the one hand $\pi^{0}\pi^{0}$, $\eta\eta$ etc. final states are to be measured in order to search for exotics, 
hybrids and glueballs. On the other hand the goal is to detect DVCS photons at largest statistics, wherefore 
also the two photons from $\pi^{0}$'s decaying need to be detected at high efficiency in order to ensure excellent 
background suppression.  
ECAL1 is $3.97 \times 2.86\,{\rm m}^{2}$ large, consists of lead glass blocks of three different sizes and has 1492 channels, 
whereas ECAL2 measures $2.44 \times 1.83\,{\rm m}^{2}$, applies three different type of modules (all of same size) and comprises 
3072 channels in total, see also \cite{spectro}. ECAL1 is in operation since 2006, and ECAL2 has partly been upgraded for 
the 2008 running: For the central part the lead glass GAMS blocks have been replaced by so-called Shashlik sampling modules 
newly developed at IHEP Protvino to cope with the higher irradiation dose and to improve the energy resolution for the small 
angle regime at high energies. In addition, most part of ECAL2 read-electronics have been upgraded (from 10 to 12 bit SADCs). 
ECAL1 provides a larger angular acceptance and detects on average photons of lower energy as
compared to ECAL2, see Figs.\,\ref{fig.dvcsEdistri} and \ref{fig.ecals1u2}. 

The minimum DVCS photon energy to be detected in
ECAL1 and ECAL2 is 5\,GeV and 10\,GeV respectively. Consequently, also $\pi^{0}$ of same energies have to be detected for 
background suppression. Since the $\pi^{0}$ decays into two photons, the lower energy threshold needed for efficient background
suppression is determined by the lower energetic photon from the decay. A kinematics calculation deliver a lower energy
threshold of less than 1.25\,GeV (0.73\,GeV) and 2.5\,GeV (1.5\,GeV) for ECAL1 and ECAL2 respectively in order to achieve a 
detection probability of better than 50\,\% (70\,\%). 
In Fig.\,\ref{fig.ecalCoverage} the calorimeter coverage of DVCS photon detection in terms of aperture (as 
foreseen for the future) is shown. 
In such a scenario\footnote{ECAL0 assumed to be 2$\times$2\,m$^{2}$ large, having a central hole 
of 1.2$\times$1.0\,m$^{2}$ and being located at about 2.5\,m downstream of the target.} we were able to 
detect $\sim\;$90\,\% of the total number of DVCS photons produced.
Nearly half of these photons, namely 43\,\%, would have to be detected by ECAL1, and about a quarter  
by ECAL2 (23\,\%) and ECAL0 (22\,\%) respectively.    
 
In conclusion, for DVCS photon detection and good $\gamma/\pi^{0}$ separation, we need lower energy thresholds of $\sim\;$1\,GeV and 
$\sim\;$2\,GeV in ECAL1 and ECAL2 respectively; since the hardware thresholds are in the order of 150\,MeV, there is, a priori, 
no limitation for lower energy thresholds below 1\,GeV.   
%
%...................................................
%
%   Section: Prfommm
%
%...................................................
\section{Measurements at CERN T9 test beam facility}
\label{sec.T9}
\vspace{-0.25cm}
A test beam campaign has been performed at the T9 PS beam line in October 2007. The project was organised by the 
Protvino IHEP Group, first of all to characterise the new COMPASS radiation hard Shashlik modules developed at IHEP, 
and also to study the present performance of the COMPASS Calorimetry for better understanding and improving the 
capability for excellent calorimetry as needed for the COMPASS Hadron and DVCS future programmes. 
%Due to the large experimental and analysis effort, a joint ``COMPASS@T9Group'' has been formed: Dubna, Freiburg, 
%Mainz, TU M\"{u}nchen, Protvino, Saclay, Torino. 

\begin{figure}[t]
\noindent
\centering
\includegraphics[clip, trim= 50 30 30 140,width=0.95\linewidth]{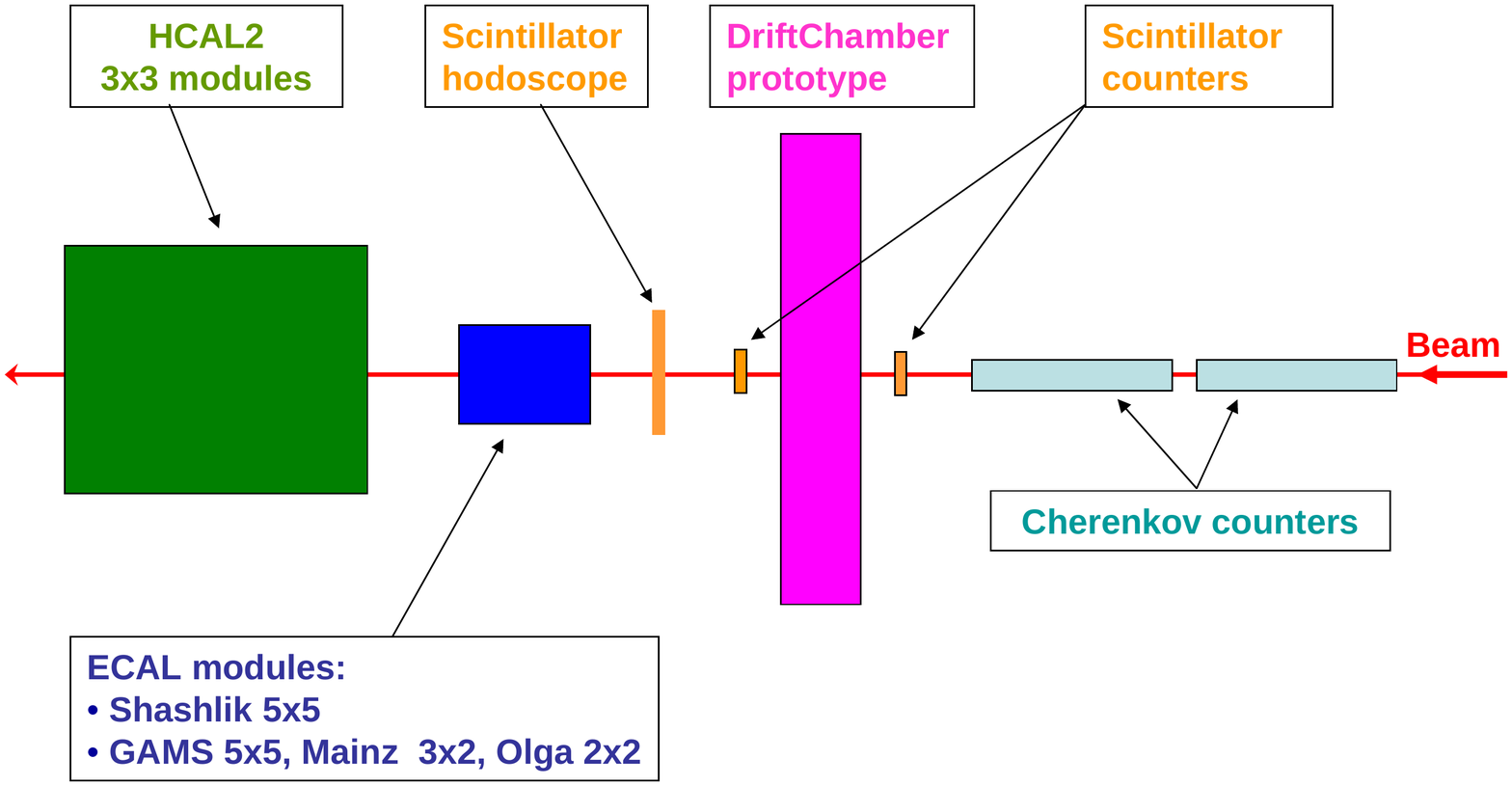}
\caption{\emph{Experimental setup at CERN T9 test beam facility.}}
\label{fig.setupT9}
\end{figure}

At the CERN PS T9, a beam containing electrons, muons, 
pions and protons is available\footnote{The beam charge is revertable at T9; due to the higher flux we used positive particles 
(electrons and muons for the studies presented).} in the energy range of 1-15\,GeV. Two Cherenkov threshold counters belong to the 
infra-structure and allow for triggering on the different particles in the beam. 
Fig.\,\ref{fig.setupT9} shows the experimental setup. Different types of COMPASS ECAL modules as well as 
3x3 HCAL modules (and a drift chamber in front, as in the spectrometer) had been installed.

Different measurements were performed. Main goals of the measurements were the determination of energy resolution and 
uniformity of the different ECAL modules (5x5 matrix of GAMS modules, 3x2 matrix for Mainz and 2x2 for Olga were used), 
especially the performance comparison between the GAMS to the newly developed Shashlik modules, which replace the
GAMS modules in the central part of ECAL2 for the 2008 run, cf. Fig.\,\ref{fig.ecals1u2}. For these studies, electrons were selected in the beam
momentum range of 1-5\,GeV, also data with muons and electrons of known energy, in the trigger were taken, cf. Sec.\,\ref{subsec.muCalib}.
Moreover, data were taken with pions in the beam momentum range of 1-10\,GeV in order to study the combined response
of electromagnetic and hadronic calorimeters.

Not all studies are yet enclosed. First results needed for the 2008 run preparation and as input for the 
DVCS physics proposal currently under work are discussed. Preliminary results of performance studies of the 
existing GAMS lead glass modules are exemplary shown in Sec.\,\ref{subsec.EcalPerf}, and investigations on how to use muon signals for calibration 
and monitoring issues are discussed in Sec.\,\ref{subsec.muCalib}. 
%
%
%-------------------------------------------------------
%
%   SubSection:GAMS performance
%
%------------------------------------------------------
\subsection{ECAL performances based on T9 test beam data}
\label{subsec.EcalPerf}
\vspace{-0.25cm}
\begin{figure}[t]
\begin{minipage}[c]{.48\linewidth}
\begin{center}
\includegraphics[clip, trim= 15 0 100 440,width=1.0\linewidth]{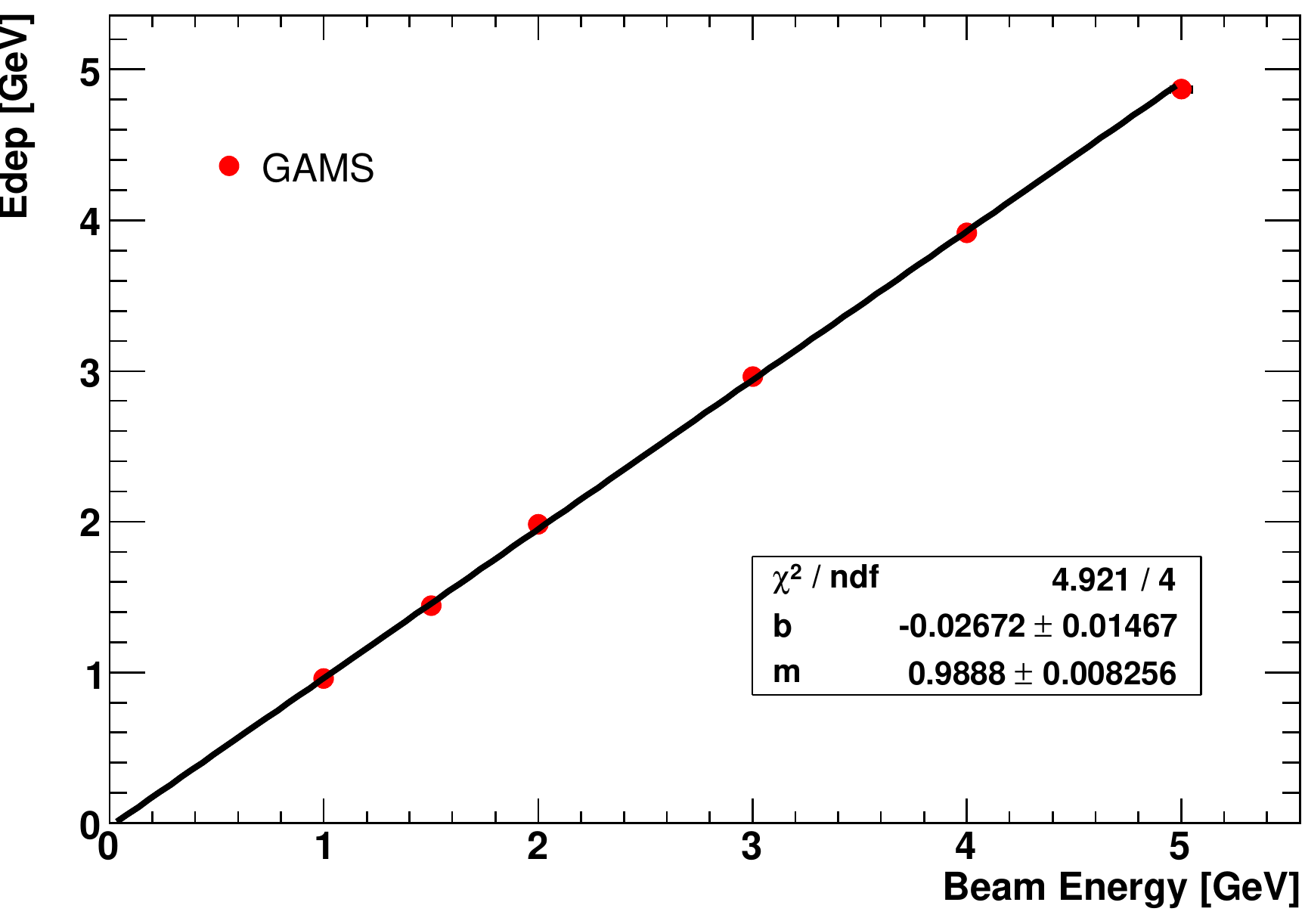}
\end{center}
\end{minipage}\hfill
\begin{minipage}[c]{.48\linewidth}
\begin{center}
\includegraphics[clip, trim= 5 0 100 430,width=1.0\linewidth]{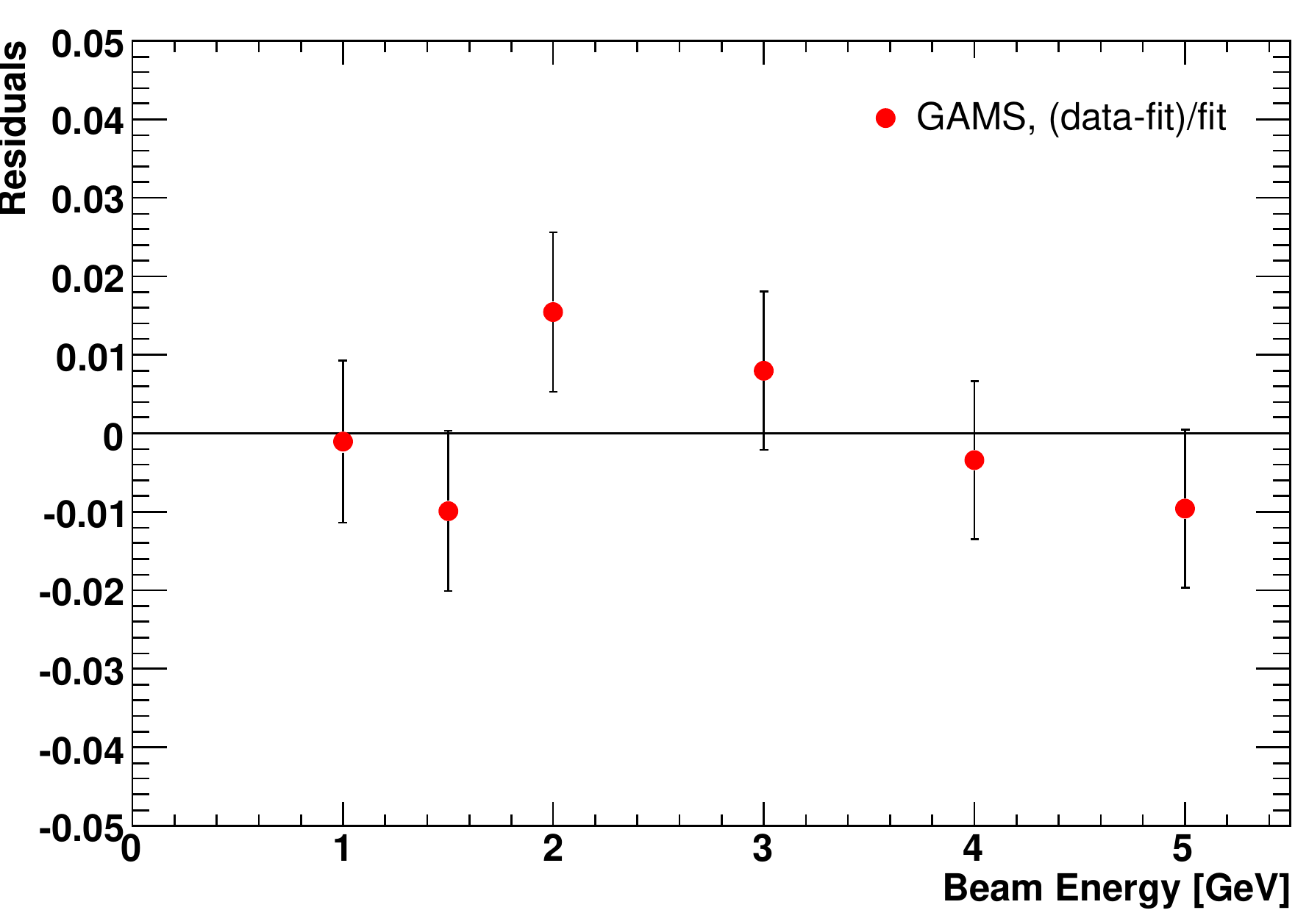}
\end{center}
\end{minipage}
\caption{\emph{Linearity of COMPASS ECAL GAMS modules as measured at T9: {\rm (left)} Measured energy deposit vs. beam energy;
{\rm (right)} Residuals of fit to the data points as shown left. (Preliminary)}}
\label{fig.T9_GAMS_linearity_residual_sum_9}
\end{figure}
For the GAMS blocks, a 5$\times$5 matrix was installed and calibrated using the T9 electron 
beam of known energy, namely 4 and $5\,$GeV respectively. Calibration involves two main steps:
\begin{enumerate}
\item[1.)] Inter-calibration between different cells with the beam centered in each cell.
\item[2.)] Global calibration by summing up the energy deposit using the surrounding cells - summation $1 + 8 = 9$ is 
done (the central one and the 8 neighbours of the one the beam is centered on.) - and the summed energy is then 
scaled to the reference, namely the beam energy. 
\end{enumerate}
The final calibration using this approach reads as $E_{{\rm dep}} = 1/C \cdot \sum c_{{\rm i}} \cdot A^{{\rm i}}_{{\rm el}}~,$
where $A^{i}_{{\rm el}}$ are the measured electron amplitudes in individual cells, $c_{i}$ the corresponding calibration coefficients 
obtained by the 1st step, the inter-calibration, and $C$ the global calibration coefficient from the 2nd step.   
The measured energy deposit $E_{\rm dep}$ summing up i=9 modules for incident electron energies ranging from 1-5\,GeV 
of the GAMS modules is exemplary shown in Fig.~\ref{fig.T9_GAMS_linearity_residual_sum_9}~-~left. The deviations from 
the linear fit: $\rm (data - fit)/fit$, is given by Fig.~\ref{fig.T9_GAMS_linearity_residual_sum_9}~-~right. The error 
bars along the ordinate take into account only the statistical error, whereas for the abscissa, an uncertainty of beam 
particle momenta of 1.0\,\% has been estimated. The relative differences from the linearity of up to ~2\,\% in the energy 
range studied are mainly within the uncertainty of the initial beam energy. The beam energy error has been incorporated 
into the statistical error on the ordinate. These differences could hint at a small but systematic effect, non-linearity 
of $1$ to $2\,\%$ have also been observed for lead glass calorimeters by other experiments at few GeV energies~\cite{caloTestPhenix}.

\noindent
\begin{wrapfigure}{l}{0.52\textwidth}
\begin{center}
\vspace{-0.9cm}
\includegraphics[clip, trim= 0 0 100 430,width=1.0\linewidth]{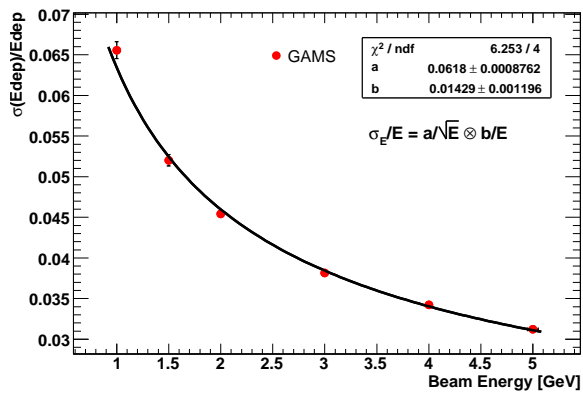}
\end{center}
\vspace{-0.4cm}
\caption{\emph{Energy resolution of GAMS lead glass modules as measured at T9 test beam facility. (Preliminary)}}
\label{fig.resolutionGAMS}
\vspace{-0.9cm}
\end{wrapfigure}
Fig.~\ref{fig.resolutionGAMS} shows the relative width of the energy deposit $\sigma(E_{\rm dep})/E_{\rm dep}$ (energy 
resolution) distribution as a function of the incident beam energy and 
the corresponding fit using two parameters. 

In conclusion, the present resolution of the GAMS modules is determined to be $\sigma_{\rm E} / E = 6.2\,\% / \sqrt{E}$ 
and a constant term of $1.4\,\%$. 
Corresponding studies for Mainz and Olga modules\footnote{Mainz and Olga modules are 
larger than GAMS by a factor of 3.85 and 14 respectively in cross-section so that summing up 3$\times$3 modules is not 
needed to minimise transversal leakage; sum over 4 modules is however needed since the beam position was close to the 
centre of 4 modules.} (not explicitly shown here) 
deliver $7.0\,\% / \sqrt{E} \oplus 1.8\,\%$ and $4.3\,\% / \sqrt{E} \oplus 3.2\,\%$ respectively, where $\oplus$ denotes 
the quadratic sum $a\oplus b = \sqrt{a^2+b^2}$. All these values have to be taken as preliminary, 
since analysis of T9 test beam data is not yet enclosed.
%
%
%-------------------------------------------------------
%
%   Section: muon caliob
%
%------------------------------------------------------
\subsection{Calibration using muons}
\label{subsec.muCalib}
\vspace{-0.25cm}
To calibrate the ECALs at place inside the spectrometer with electrons necessitates a special tuning of our beam 
line and also to move the ECAL structure across the beam line. Certain limitations are inherent to this method and 
alternative options, which allow to inter-calibrate the ECAL cells and also to monitor the gain stability, are 
exploited. Hadrons and muons, which deposit a non negligible amount of energy in the ECALs, can be used to complement 
calibration, as done in other experiments~\cite{cmsInterCalib}. 

A cross-calibration using electrons and muons has been performed on the T9 test beam data reported 
here. The muon signal results from all the photons detected, which have two main sources: Cherenkov light 
produced directly by the muon (well above threshold) and Cherenkov light produced indirectly by the 
ionization losses processes (e.g. $\delta$-electrons). Precise prediction of the muon signal amplitude in 
equivalent electromagnetic energy requires a full simulation (under way). For the cross-calibration technique, however, 
such complex task is, a priori, not essential since this method involves measuring the ratio $A_{\rm el}/A_{\rm \mu}$ 
between the muon and electron signal amplitudes. 
\begin{figure}[t]
\begin{minipage}[c]{.38\linewidth}
\begin{center} 
\includegraphics[clip,angle=90, trim= 5 45 54 30,width=1.0\linewidth]{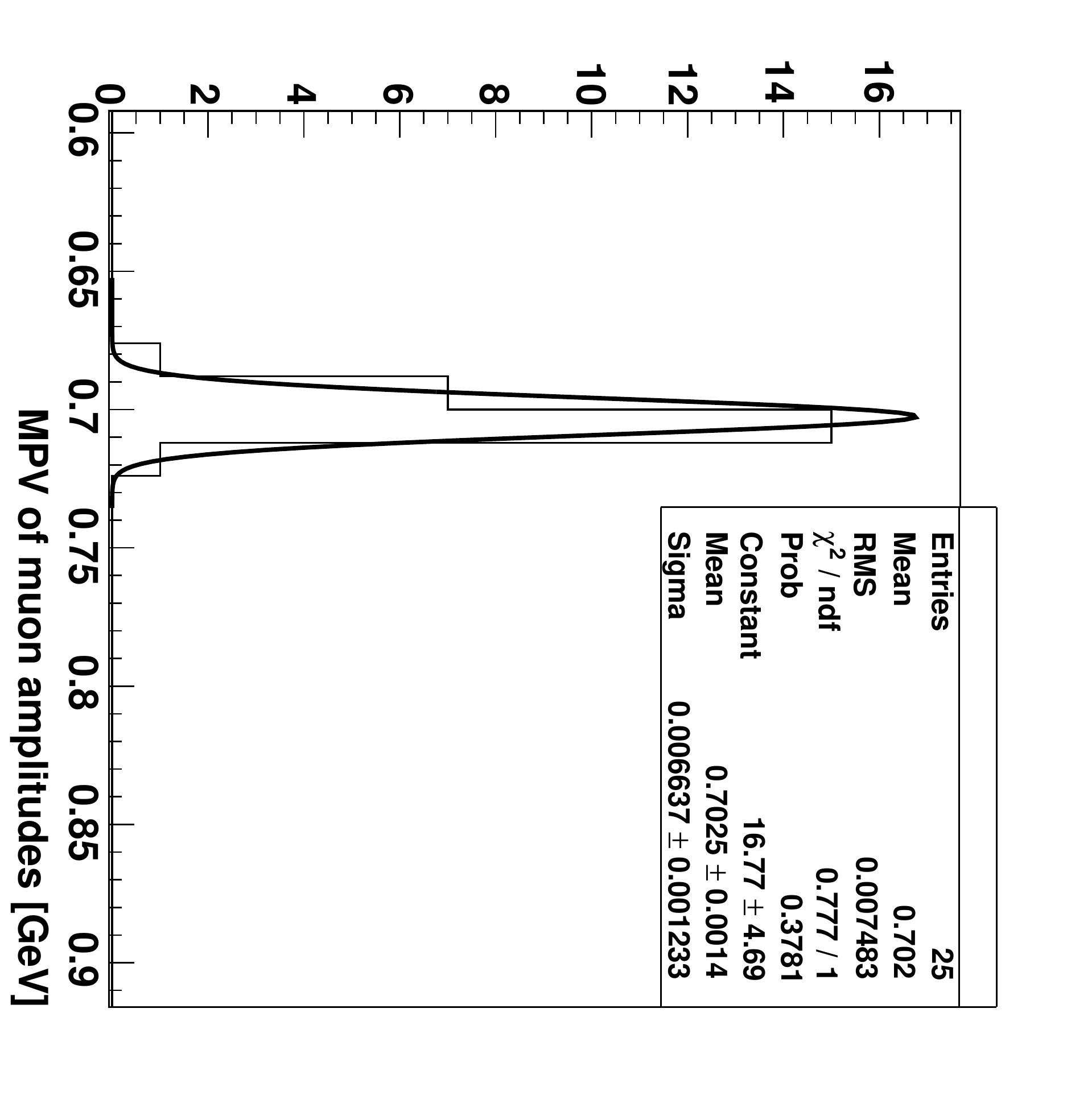}
\end{center}
\end{minipage}\hfill
\begin{minipage}[c]{.58\linewidth}
\begin{center}
\includegraphics[clip, trim= 10 3 100 440,width=1.0\linewidth]{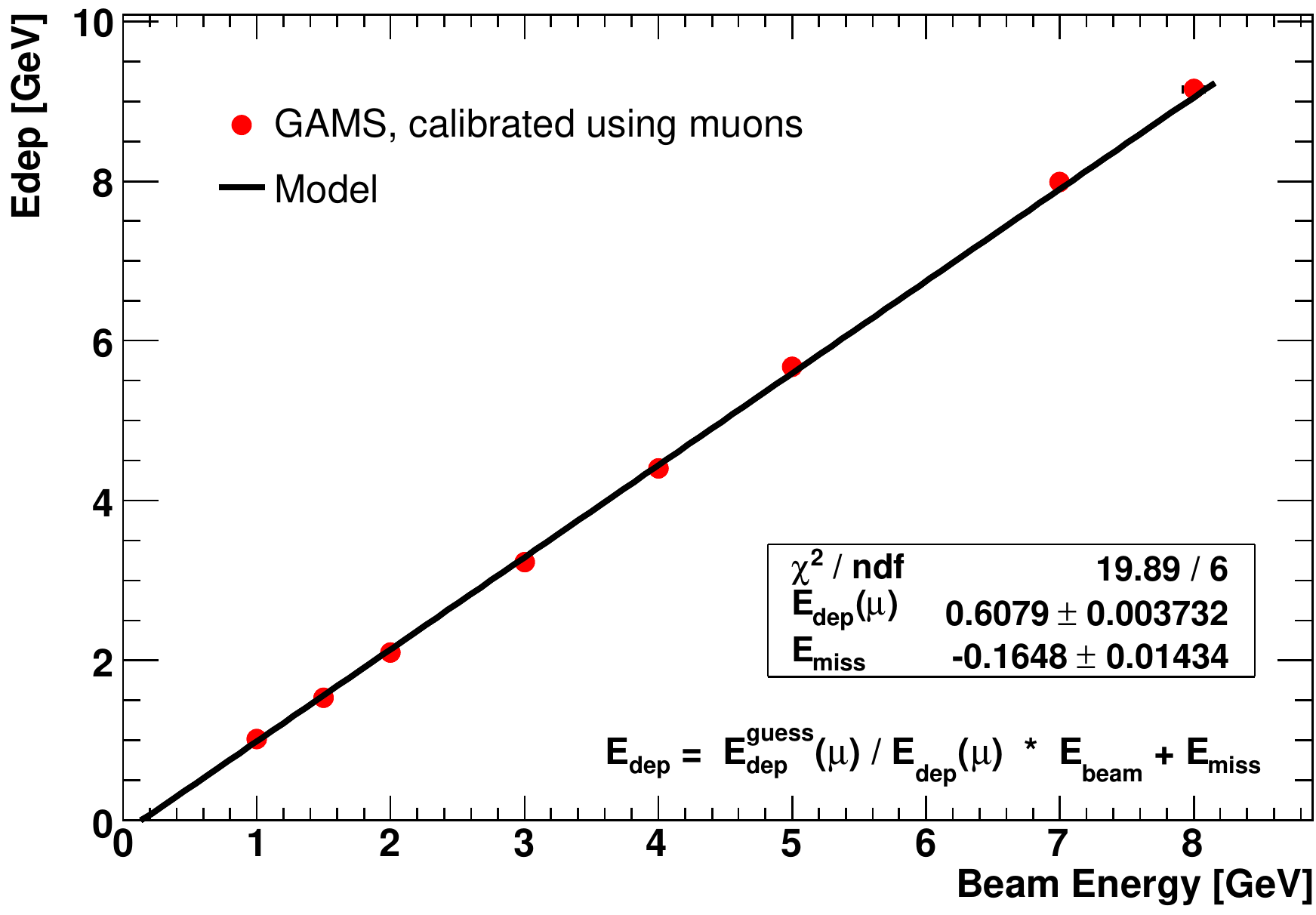}
\end{center}
\end{minipage}
\caption{\emph{{\rm (left)} Distribution of fitted muon amplitudes after inter-calibration using muon signals, 
mean value $A_{\rm \mu}=0.703~(\sigma = 0.007)$\,GeV. {\rm (right)} A linear fit to $E_{\rm dep}$ values obtained 
for GAMS for T9 electron beam energies ranging from 1 to 8\,GeV. Fitted parameters are $E_{\rm dep}(\mu)=0.608\pm0.004$\,GeV
and $E_{\rm miss}=-0.165\pm0.014$\,GeV.}}
\label{fig.muonCalib_a}
\vspace{-0.2cm}
\end{figure}

The $5~$GeV muon signal in the GAMS cells was assumed, as a first guess to correspond to 
$E_{\rm dep}(\mu) = 0.7$\,GeV providing the 1st calibration (better than 1\,\%), see Fig\,\ref{fig.muonCalib_a} - left, 
which can in principle be done with any value, cf. Sec.~\ref{subsec.EcalPerf}. 
The T9 electron beam energy was varied from 1-8\,GeV and the corresponding energy deposit $E_{\rm dep}$
was measured, summing up 9 modules and using this 1st calibration. It was previously established, see 
Fig.~\ref{fig.T9_GAMS_linearity_residual_sum_9}, that the GAMS cells have excellent linear response to 
electrons in this energy range. A linear fit to the full set of $E_{\rm dep}$ measured after this 
inter-calibration was performed with two parameters: $E_{\rm dep}(\mu)$ corresponding to the true effective 
muon energy deposit and $E_{\rm miss}$ representing the minimum measurable $E_{\rm dep}$ (lower energy threshold). 
Result from the fit is shown in Fig.~\ref{fig.muonCalib_a} - right. One sees that the first guess of $E_{\rm dep}(\mu)=0.7~$GeV leads 
to an overestimate of electron energy by $\sim$15\,\% at e.g. 5\,GeV.   
As a cross-check, the different cells were inter-calibrated using the value of $E_{\rm dep}(\mu)$ from this 
fit and the same quality in describing the data is achieved when just fitting the one parameter $E_{\rm
miss}$, i.e. the overestimation of electron energies is corrected when inter-calibrating the individual cells 
directly to $E_{\rm dep}(\mu)=0.608$\,GeV, namely the correct value as indirectly measured or obtained in 
Fig.\,\ref{fig.muonCalib_a} - right. 
\begin{figure}[t]
\begin{minipage}[c]{.38\linewidth}
\begin{center} 
\includegraphics[clip, trim= 30 12 90 315,width=1.0\linewidth]{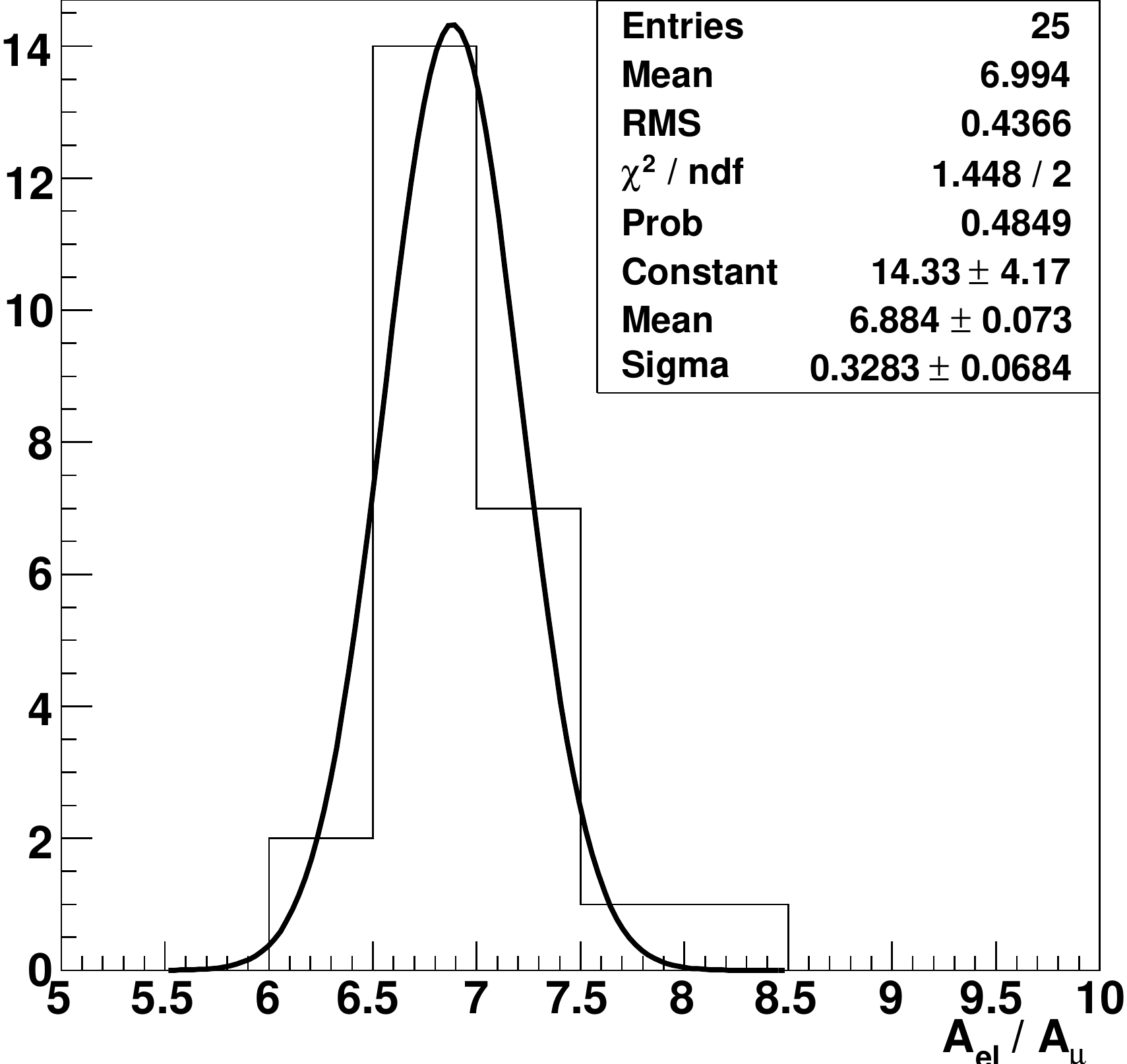}
\end{center}
\end{minipage}\hfill
\begin{minipage}[c]{.58\linewidth}
\begin{center}
\includegraphics[clip, trim= 135 125 180 110,width=1.0\linewidth]{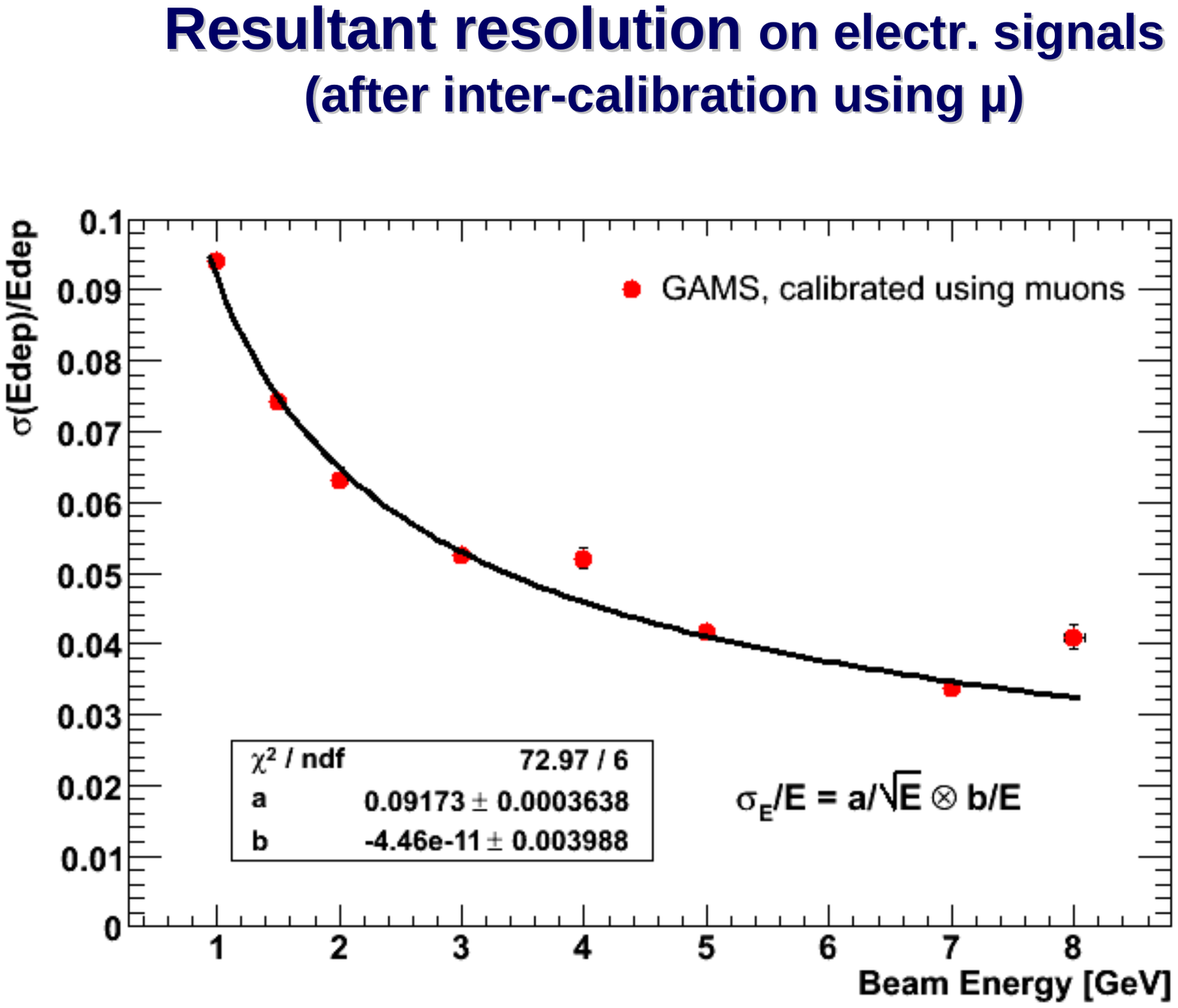}
\end{center}
\end{minipage}
\caption{\emph{{\rm (left)} Distribution of $A_{\rm el}$ to $A_{\rm mu}$ ratio of GAMS cells, 
mean value $A_{\rm el} / A_{\rm\mu}=0.688~(\sigma = 0.328)$. {\rm (right)} Energy resolution of GAMS lead glass modules obtained using muon signals. The ECAL GAMS modules have been 
inter-calibrated with muon signals and cross-calibrated to electron signals of known energies as measured at T9 test beam 
facility, see also Fig.\,\ref{fig.muonCalib_a}. (Preliminary)}}
\label{fig.muonCalib_b}
\vspace{-0.2cm}
\end{figure}

The resulting energy resolution $\sigma(E_{\rm dep})/(E_{\rm dep})$ as a function of electron beam energy is 
given in Fig.~\ref{fig.muonCalib_b} - right. These results based on the inter-calibration of GAMS cells 
using muon signal and a cross-calibration to electron signals of known energy should be compared to the 
similar results shown in Figs.\,\ref{fig.T9_GAMS_linearity_residual_sum_9} and \ref{fig.resolutionGAMS} obtained 
doing the full procedure using electrons only. The resolution obtained with this method is worse by 
merely $\sim\;$2-3\,\%, which is consistent with the spread observed in the distribution of $A_{\rm el}/A_{\rm \mu}$ 
ratios of individual cells, given in Fig.~\ref{fig.muonCalib_b} - left. 
Indeed the spread for all 25 GAMS cells is about 5\,\%, however, the central GAMS cell contains about 85\,\% of total 
electron beam energy deposit (in the given range of few GeV), and the central one was checked to deviate by $\sim$\,2.5\,\% 
from the mean value of all 25 $A_{\rm el}/A_{\rm \mu}$ ratios.
The further 8 cells involved in the $E_{\rm dep}$ measurement show a spread of $\sim$\,5\,\%. Weighting these inaccuracies 
correspondingly leads to a decrease in resolution by 2-3\,\%. 
%-------------------------------------------------------
%
%   Section: Conclusions
%
%------------------------------------------------------
\section{Conclusions}
\label{sec.7}
\vspace{-0.25cm}
Excellent COMPASS calorimetry is mandatory for both, the present 2008/09 COMPASS Hadron running as well 
as the GPD measurements via DVCS as foreseen at COMPASS for the future.
The present performances of existing ECAL lead glass blocks (GAMS, Mainz and Olga) have been quantified, 
and the derived numbers serve as realistic input for ongoing Monte Carlo Simulations. It should be noted that 
a simplified method for calibration was applied here as compared to the more sophisticated procedure within the official 
COMPASS reconstruction (of much larger number of ECAL channels) comprising multiple iterations. 
It has been shown that for calibration and monitoring issues, the signals from muons can be used for
inter-calibration. Together with a cross-calibration to electron signals of known energy, this calibration
procedure turns out to be merely worse by 2-3\,\% as compared to the full calibration using electrons. A
proof of principle has been provided for the GAMS cells.   
%-------------------------------------------------------
%
%   Section:Acknowledge
%
%------------------------------------------------------
\section{Acknowledgements}
\label{acknow}
\vspace{-0.25cm}
The corresponding author acknowledges travel support by the BMBF (Germany). 
%
% BibTeX users please use
% \bibliographystyle{}
% \bibliography{}
%
% Non-BibTeX users please use

\end{document}